\documentclass[aps, prd, twocolumn, nofootinbib]{revtex4}

\usepackage{amsmath, amssymb, bm}
\usepackage{graphicx}
\usepackage{epstopdf}
\usepackage{amsfonts}
\usepackage{amssymb}
\usepackage{amsbsy}
\usepackage{amsmath}
\usepackage{latexsym}
\usepackage{sansmath}
\usepackage{subfigure}
\usepackage{lipsum}
\usepackage{float}
\usepackage{cancel}

\usepackage{bm}
\usepackage{color}

\usepackage{hyperref}

\usepackage{comment}
\usepackage{csquotes}
\usepackage{physics}
\usepackage{accents}

\def\bea {\begin{eqnarray}}
\def\eea {\end{eqnarray}}

\def\lp{\ell_{\rm Pl}}

\begin{document}

\title{Testing ER = EPR with hydrogen}

\author{Irfan Javed} \email{i.javed@unb.ca}
\affiliation{Department of Mathematics and Statistics, University of New Brunswick, Fredericton, Canada}

\author{Edward Wilson-Ewing} \email{edward.wilson-ewing@unb.ca}
\affiliation{Department of Mathematics and Statistics, University of New Brunswick, Fredericton, Canada}

\begin{abstract}

According to the ER = EPR conjecture, entangled particles are connected by quantum wormholes. Under the assumption that some of the electric field surrounding an entangled charged particle leaks into the wormhole, we show that this effect will modify the hyperfine structure of the hydrogen atom. In addition, if the quantum wormholes are non-traversable, this will also lead to a non-zero total effective charge for the hydrogen atom. These effects provide strong constraints on the amplitude of this potential ER = EPR effect, given high-precision measurements of the hydrogen atom's hyperfine structure and total charge.

\end{abstract}

\maketitle

\textbf{\textit{Introduction}}---Recent research in quantum gravity indicates that entanglement and spacetime connectivity are closely related, with entanglement perhaps even generating connectivity. For example, in the duality between asymptotically anti-de Sitter spacetimes and conformal field theories (AdS/CFT), the spacetime connectivity in the AdS theory is related to entanglement entropy in its dual CFT \cite{Ryu:2006bv, Ryu:2006ef, VanRaamsdonk:2010pw, Lashkari:2013koa, Faulkner:2013ica}. In loop quantum gravity, on the other hand, there is entanglement entropy between connected nodes in spin-networks \cite{Husain:1998zw, Donnelly:2008vx}, and quanta of geometry can be connected to each other by entangling the quantum degrees of freedom lying on the surfaces that are to be glued together \cite{Oriti:2013aqa, Chirco:2017vhs, Baytas:2018wjd}. Motivated in part by these results, it has been suggested that the microscopic connectivity of spacetime might be associated with entanglement entropy in any theory of quantum gravity \cite{Bianchi:2012ev, Swingle:2014uza}. If this is indeed correct, then it may also be possible to entangle geometric degrees of freedom located in different spacetime regions and generate a quantum wormhole bridging them (as has been previously proposed in other contexts \cite{Smolin:1994uz, Markopoulou:2007ha, Konopka:2008hp, Prescod-Weinstein:2009bqa, Hossenfelder:2013yda}).

A stronger conjecture has since been put forward by Maldacena and Susskind who argued that, more generally, entangled degrees of freedom (not just quanta of geometry or their CFT duals as considered above) are connected by a quantum wormhole, including simple cases like the singlet state of two spins \cite{Maldacena:2013xja}. This proposal for wormhole-mediated entanglement is commonly called ER = EPR, its name stemming from how it suggests that quantum Einstein-Rosen bridges (or wormholes) accompany Einstein-Podolsky-Rosen pairs (or entangled particles); it is important that the wormholes be quantum in order to properly capture tripartite entanglement \cite{Gharibyan:2013aha, Balasubramanian:2014hda}. Further, since entanglement is the cause of the wormhole, destroying the entanglement---say, through a measurement---would result in the disappearance of the wormhole linking the electrons. Note that this conjecture is different from AdS/CFT, where entanglement in the boundary CFT generates connectivity in the spacetime of the dual AdS theory; rather, in ER = EPR, the wormholes connect different regions of the same spacetime that the EPR entangled pairs live in \cite{Susskind:2017ney}. This is also different from loop quantum gravity, where connectivity is created specifically by the entanglement of quanta of geometry, but not through the entanglement of other types of degrees of freedom.

In this paper, we look at some consequences of ER = EPR, focusing on the hydrogen atom. One potential effect of ER = EPR in systems of two charged particles (like hydrogen) that are entangled is on the electric field surrounding the particles: if a wormhole is present in the vicinity of an entangled charged particle, some of its electric field could leak into the wormhole \cite{Dai:2019mse, Wilson-Ewing:2021kcp}. If this occurs, then as a consequence the strength of the entangled particle's electrostatic interaction with other particles in its surroundings will be reduced, in turn modifying the energy spectrum of the entangled system's Hamiltonian. (For other potential tests for ER = EPR, see \cite{Dai:2020ffw}.)

In the following, we focus on nontraversable wormholes, as suggested in the original version of ER = EPR  \cite{Maldacena:2013xja}.  Nonetheless, the case of traversable wormholes is also interesting (it has recently been pointed out that traversable worm holes can be created without exotic matter fields \cite{Gao:2016bin, Blazquez-Salcedo:2020czn}), and we will briefly discuss this possibility in the Appendix.

\textbf{\textit{Basic assumptions}}---To make concrete calculations possible, it is necessary to make some basic assumptions concerning how the ER = EPR conjecture is realized in systems of two particles.

The key assumption of this paper, following \cite{Dai:2019mse, Wilson-Ewing:2021kcp}, is that part of the electric field surrounding an entangled charged particle will enter the wormhole, thereby reducing the electric field surrounding the entangled particle as measured by an observer who does not have access to the wormhole. We further assume that this effect is independent of the particle's motion, or its possible quantum superposition in different positions.

In addition to this key assumption, we make three further assumptions. First, we assume that the fraction of the electric field that enters the wormhole is proportional to the entanglement entropy $s$ between the entangled pair; this is motivated by the proposal that the cross-sectional area of such a wormhole should be proportional to the entanglement entropy \cite{Verlinde:2020upt} (for an alternate proposal that the area of the throat of the wormhole should be zero, see \cite{Jusufi:2025rlr}; in this case, there would be no effect), and that in the limiting case of no entanglement, there would be no wormhole and no effect. Second, we assume that this effect impacts point particles, but not composite objects (like protons, for example) whose size is much larger than $\lp$, the scale of quantum gravity. Third, we assume that electrons are point particles, and (when entangled) some of the electric field surrounding them will pass into the ER = EPR wormhole.

The assumption that this effect is important for point particles but not composite objects can be understood to require that the mouth of the wormhole must be localized, and will necessarily be near the charge carried by a point particle, but will typically be at a large distance (compared to the Planck length) from the charge-carrying point particles within a composite particle like a proton. As stated above, we also assume that the possible quantum superposition of a particle in different positions has no impact on the strength of this effect---this is important because the wave function for a point particle will typically be a wave packet. To see why this last assumption is reasonable, consider an (entangled) electron passing through an interferometer, such that it is in a superposition of being in two locations; then, this assumption corresponds to the requirement that the mouth of the ER = EPR quantum wormhole not be located halfway between those two locations, but rather be in the same superposition as the electron. The same argument applies to a wave packet, or an electron in an orbital: the state for the location of the mouth of the wormhole is in the same superposition of positions as the particle itself. This is why it is reasonable to assume that the particle being delocalized has no impact on this effect.

With these assumptions, the Gauss law for a surface surrounding an entangled (point) particle of charge $q_e$ has the form
\begin{equation}
\frac{q_{e}}{\epsilon_{0}} = \oint_{S}\mathbf{E}_{S}\cdot d\mathbf{a}+\oint_{W}\mathbf{E}_{W}\cdot d\mathbf{a},
\label{eq. 1}
\end{equation}
where it is necessary to include a contribution from the wormhole. Assuming the surface is a sphere at distance $r$ from the entangled particle, we parametrize the loss of the electric field through the wormhole by expressing the resulting electric field on this surface $S$ as
\begin{equation}
\mathbf{E}_{S} = \frac{q_{e}/4\pi\epsilon_{0}r^{2}}{1+s/\pi\alpha^{2}} \, \hat{\mathbf{r}},
\label{eq. 2}
\end{equation}
where $\hat{\mathbf{r}}$ is the radial unit vector pointing away from the electron. Due to the loss of electric flux into the wormhole, the charge of an entangled electron is effectively suppressed by a factor of $1+s/\pi\alpha^2$, where $s$ is the entanglement entropy between the entangled pair and $\alpha$ is a dimensionless parameter that determines the strength of this effect, with $\alpha\to\infty$ corresponding to the case that no electric field at all enters the wormhole. (Note that while the specific form of the parametrization is motivated by semi-classical arguments \cite{Wilson-Ewing:2021kcp}, this parametrization is nonetheless fully general.) Semi-classical arguments lead to the expectation that $\alpha \sim \mathcal{O}(1)$, as do naturalness arguments that dimensionless parameters should typically be of order 1; nonetheless, we keep $\alpha$ fully general here.

\textbf{\textit{Hydrogen}}---Now, consider a hydrogen atom where the proton and the electron are entangled, so there will be a quantum wormhole connecting the two according to the ER = EPR conjecture. Note that hydrogen possesses an intrinsic entanglement \cite{Tommasini:1997pm, Qvarfort:2020qgl}, since the energy eigenstates of the hydrogen atom are products of the state of its center of mass and the state of the separation vector from the proton to the electron, the latter denoted by the usual $\ket{n, \ell, m}$ state. In these product states, however, the centers of mass of the proton and electron are entangled. On top of this, there could be some additional entanglement between the two, for example, through their spins. For the sake of simplicity, in the following, we mostly focus on entanglement between the spins of the proton and electron.

Given our assumptions above, we can ask how much of the electric field passes into the wormhole. This question is interesting since the hydrogen atom is perhaps the most well-studied bipartite system in physics. In particular, its energy spectrum has been used to test quantum theories for over a century since it can be calculated very precisely from first principles and measured very accurately \cite{Mohr:2024kco}, with the energies known accurately to $12$ significant figures \cite{Grinin:2020txk}; techniques like high-resolution laser spectroscopy yield even more accurate measurements, with some transitions having been measured with an accuracy of up to $15$ significant figures \cite{Mohr:2024kco}. This accuracy can be used to test ER = EPR.

As discussed above, we assume the wormhole has no effect on the proton since the proton's radius is much larger than the Planck length. On the other hand, we assume the electron is a point particle \cite{Bourilkov:2001pe, Gabrielse:2006gg} and therefore that some of the electric field surrounding the electron does go into the wormhole, thereby reducing the electric field surrounding the electron following \eqref{eq. 2}. (Of course, the electron in a hydrogen atom is in a superposition of positions, but as stated above, we assume this has no impact on the magnitude of the electric field that leaks into the wormhole.)

Given this, to calculate the electric potential energy appearing in the Schr\"odinger equation for the hydrogen atom in the presence of entanglement, we should use the usual expression for the electric field due to the proton $\mathbf{E}_{p}$ but use \eqref{eq. 2} for the electric field from the electron with the result that (discarding, as usual, the self-energy contributions)
\begin{equation}
V_{\text{int}} = \epsilon_0\int d^3x \, \mathbf{E}_{p}\cdot\mathbf{E}_{e}
\end{equation}
is reduced by a factor of $1+s/\pi\alpha^{2}$ due to the modification to $\mathbf{E}_{e}$ as given in \eqref{eq. 2}; this can be viewed as an effective suppression of the electric charge of the electron from $q_{e} = -e$ to $q_{e}' = -e' = -e/(1+s/\pi\alpha^{2})$.

Importantly, this effect suppresses the energy eigenvalues of the hydrogen atom by two powers of the same factor, with the energy becoming
\begin{equation}
\begin{aligned}[b]
E_{n} =&\:-\frac{m_{e}}{2\hbar^{2}}\left(\frac{ee'}{4\pi\epsilon_{0}}\right)^{2}\frac{1}{n^{2}}\\
=&\:-\frac{m_{e}}{2\hbar^{2}}\left(\frac{e^{2}}{4\pi\epsilon_{0}}\right)^{2}\left(1+\frac{s}{\pi\alpha^{2}}\right)^{-2}\frac{1}{n^{2}},
\end{aligned}
\label{eq. 4}
\end{equation}
where $m_{e}$ is the mass of the electron, $e$ is the charge carried by the proton, and $n$ is the principal quantum number. A convenient way to test this modification to the energy spectrum is by examining the hydrogen atom's hyperfine structure.

\textbf{\textit{Hyperfine structure}}---The hyperfine structure of hydrogen depends on the spin of the proton and the electron, with a split between the three higher-energy states that have a total spin of $j = 1$ (namely, the states $\ket{\uparrow\uparrow}$, $\tfrac{1}{\sqrt2}(\ket{\uparrow\downarrow}+\ket{\downarrow\uparrow})$, and $\ket{\downarrow\downarrow}$) and the lower-energy singlet state $\tfrac{1}{\sqrt2}(\ket{\uparrow\downarrow}-\ket{\downarrow\uparrow})$ with total spin $j = 0$. Here, as usual, we take the principal quantum number to be $n = 1$ for the discussion of hyperfine structure. This split is due to the electron being in the magnetic field of the proton, and the magnitude of the split depends on the strength of the magnetic field felt by the electron.

If ER = EPR holds and some of the electric field is lost into the wormhole, an entangled electron will be further from the proton than in the unentangled case. This is because we can roughly quantify the atomic size of the hydrogen atom via the Bohr radius $a_{0} = 4\pi\epsilon_{0}\hbar^{2}/(m_{e}e^{2})$, and when the proton and electron are entangled then (because the electron's charge is effectively $e'$), this radius becomes $4\pi\epsilon_{0}\hbar^{2}/(m_{e}ee') = a_{0}(1+s/\pi\alpha^{2})$, and the atomic size of hydrogen increases.

The hyperfine splitting with ER = EPR then must be smaller since the electron---being farther away---will feel a diminished magnetic field. In precise terms, the hyperfine split without ER = EPR is \cite{griffiths2018introduction}
\begin{equation}
\Delta E_{\text{hf}} = \frac{\hbar^{2}g_{p}e^{2}}{3\pi\epsilon_{0}m_{p}m_{e}c^{2}a_{0}^{3}},
\label{eq. 5}
\end{equation}
where $g_{p}$ is the so-called $g$-factor of the proton and $m_{p}$ is its mass. This means that in ER = EPR, $a_{0}$ is larger by a factor of $1+s/\pi\alpha^{2}$, and the hyperfine split is smaller by a factor of $(1+s/\pi\alpha^{2})^{3}$ for entangled states. Now, of the four states referred to above, two are product states ($\ket{\uparrow\uparrow}$ and $\ket{\downarrow\downarrow}$) and therefore unentangled (at least in terms of their spins) while the other two states with magnetic quantum number $m=0$, namely $\tfrac{1}{\sqrt2} (\ket{\uparrow\downarrow}+\ket{\downarrow\uparrow})$ and the $j=0$ singlet state $\tfrac{1}{\sqrt2}(\ket{\uparrow\downarrow}-\ket{\downarrow\uparrow})$, have spins that are maximally entangled with entanglement entropy $s = \ln 2$.

The standard quantum mechanics calculation for transition energy from the singlet state $\ket{j = 0, m = 0}$ to the (entangled) triplet state $\ket{j = 1, m = 0}$ is given by \eqref{eq. 5}, which as explained above is increased by a factor of $(1+s/\pi\alpha^{2})^{3}$ if some of the electric field surrounding the entangled electron leaks into an ER = EPR wormhole. This transition corresponds to the well-known $21$-cm line that has been used to map neutral hydrogen in our galaxy and to determine its spiral shape \cite{Pritchard:2011xb}, and since the wavelength of the $21$-cm line is known to $12$ significant figures \cite{hellwig2007measurement}, this constrains the parameter $\alpha$:
\begin{equation}
\left(1+\frac{s}{\pi\alpha^{2}}\right)^{3}-1 \lesssim O(10^{-12}) \implies \alpha \gtrsim O(10^{6}).
\label{eq. 5a}
\end{equation}
This is a strong constraint on $\alpha$ that in particular completely rules out $\alpha \sim O(1)$, which may have been na\"ively expected.

Since ER = EPR effects are due to entanglement, they only modify the energies of the $\ket{j = 0, m = 0}$ and $\ket{j = 1, m = 0}$ states, and for this reason, the transition from the $\ket{j = 1, m = 1}$ or $\ket{j = 1, m = -1}$ to the $\ket{j = 0, m = 0}$ state is modified in a different manner, leading to a splitting of the hyperfine structure itself that has not been observed. The hyperfine energy difference \eqref{eq. 5} may be rewritten as \cite{griffiths2018introduction}
\begin{equation}
\Delta E_{\text{hf}} = \frac{\hbar^{2}g_{p}e^{2}}{3\pi\epsilon_{0}m_{p}m_{e}c^{2}a_{0}^{3}}\left(\frac{1}{4}-\left(-\frac{3}{4}\right)\right),
\label{eq. 6}
\end{equation}
where the $1/4$ corresponds to triplet states and the $-3/4$ to the singlet state. Since ER = EPR modifies only entangled states, the energy of transition from the singlet state to one of the unentangled triplet states is then
\begin{equation}
\Delta E_{\text{hf}}' = \frac{\hbar^{2}g_{p}e^{2}}{3\pi\epsilon_{0}m_{p}m_{e}c^{2}a_{0}^{3}}\left(\frac{1}{4} + \frac{3}{4}\left(1+\frac{s}{\pi\alpha^{2}}\right)^{-3}\right),
\label{eq. 7}
\end{equation}
a transition for which the wavelength of the emitted photon differs from the other transition (from the singlet to the $|j=1,m=0\rangle$ state) only by $\Delta\lambda \sim 10^{-12}$~cm, assuming $\alpha \sim O(10^6)$. This (very small) splitting of the 21~cm line is a new effect due to the loss of the electric field into an ER = EPR wormhole.

Finally, there is also the possibility of a transition from the entangled triplet state to one of the unentangled triplet states.  The calculation is analogous to the previous one, giving
\begin{equation}
\Delta E_{\text{hf}}'' = \frac{\hbar^{2}g_{p}e^{2}}{3\pi\epsilon_{0}m_{p}m_{e}c^{2}a_{0}^{3}}\left(\frac{1}{4}-\frac{1}{4}\left(1+\frac{s}{\pi\alpha^{2}}\right)^{-3}\right),
\label{eq. 7a}
\end{equation}
which corresponds to $\lambda \sim 10^{16}$~cm for $\alpha \sim O(10^{6})$. This is another new transition that can only occur if ER = EPR breaks the degeneracy between the energies of the entangled and unentangled $j=1$ states.

\textbf{\textit{Effective total charge}}---Another strong observational constraint is provided by the observed neutrality of the hydrogen atom.
Under the assumption that the proton (being much larger than the quantum gravity scale) does not lose any of its electric field into the wormhole, but that the electron (assumed to be a point particle) does, if the ER = EPR wormhole is nontraversable, then the effective total charge of the hydrogen atom will not vanish when the proton and electron are entangled, due to the loss of some of the electron's electric field into the wormhole. It has been established to a great degree of accuracy that the charge of a proton is the same but opposite to that of an electron \cite{Mohr:2024kco}, and similarly that hydrogen atoms are electrically neutral \cite{King:1960zz} with the result that (the absolute value of) their net charge is constrained to be $\lesssim O(10^{-20})e$, which in turn constrains $\alpha \gtrsim O(10^{9})$ for $s \sim O(1)$, an even stronger constraint than the one obtained from modifications to hyperfine splitting.

Recall that hydrogen possesses an intrinsic entanglement between the centers of mass of the electron and proton \cite{Tommasini:1997pm, Qvarfort:2020qgl}. The entanglement entropy in this case is in fact divergent, so it either must be appropriately regularized or, what is typically done, a different measure of entanglement can be used. Still, despite the difficulty in quantifying this entanglement in terms of $s$, entanglement is clearly present and all hydrogen atoms should show this effect to some level, and not only those with some entanglement between their spins. Further, this intrinsic entanglement is likely at least of order unity; this expectation is supported by alternate measures of entanglement that are of order unity or greater \cite{Tommasini:1997pm, Qvarfort:2020qgl}, as well as numerical simulations of a system consisting of an electron and a proton, each trapped in a potential well (analogous to the experimental setup of \cite{Poddubny:2024jkq, Winkler:2024fjn}), that have a well-defined entanglement entropy that can be $s \sim O(1)$ even if the electron and proton are much farther apart than in an atom (in which case they have a much weaker electrostatic interaction).

\textbf{\textit{Discussion}}---In summary, the ER = EPR conjecture proposes that entangled degrees of freedom (including the important example of the singlet state of two spins) are connected by a quantum wormhole. It seems reasonable to expect that at least some of the electric field surrounding an entangled charged (point) particle will leak into such a quantum wormhole, but as shown here, the hydrogen atom provides very strong constraints on this potential effect.

Importantly, this effect is distinct from the infinite series of corrections that the hydrogen spectrum receives from perturbative quantum gravity. This can be seen quite directly for the modification to the hyperfine structure which, among other effects, breaks the energy degeneracy between the triplet states; however, perturbative quantum gravity is only sensitive to energy, and in particular is insensitive to spin degrees of freedom. For this reason, it cannot break an energy degeneracy between two different states that only differ in their spins (and, neglecting ER = EPR effects, do not differ in their energy). The same is true for the effective total charge of the hydrogen atom: perturbative quantum gravity effects cannot change the total charge of a hydrogen atom. Therefore, these effects are distinct, and can be isolated, from perturbative quantum gravity corrections.

It is worthwhile to review the assumptions underlying this result. In addition to the basic assumption that some of the electric field of an entangled charged particle will leak into an ER = EPR wormhole, and that this effect is independent of the particle's motion as well as of any possible quantum superposition in different positions, there are three further assumptions: first, that the total electric field ``lost" to the wormhole is proportional to the entanglement entropy; second, that this affects point particles but not larger composite objects whose size is much greater than the Planck scale; and third, that electrons are point particles. The first two assumptions are rather weak: since we are only calculating the amplitude of the effect when $s \sim O(1)$, we in fact do not use the assumed linear dependence of the effect on $s$; and if the electric field surrounding a proton would also leak into the ER = EPR wormhole, then it can be checked that the effect would be even stronger.  The main assumption seems to be the third one, namely that the electron is a point particle. While there is so far no evidence that the electron might not be point-like, the upper bound on the size of an electron \cite{Bourilkov:2001pe, Gabrielse:2006gg} remains far from the Planck scale. (If it turns out that the electron is not a point particle, but is rather a composite object with a radius $r_e \sim 10^n \lp$ for some $n$, then the amplitude of the effect will change, depending on the microscopic physics underlying ER = EPR. For example, for the particular scenario proposed in \cite{Wilson-Ewing:2021kcp}, the constraints on $\alpha$ would weaken by approximately $n$ orders of magnitude: if $r_e \sim 10^2 \lp$ then $\alpha \gtrsim 10^7$ rather than $\alpha \gtrsim 10^9$ as derived for the point particle case, while if $r_e \gtrsim 10^9 \lp$ then $\alpha \sim O(1)$ is allowed.) We also assumed the wormhole is non-traversable, but the calculation for a traversable wormhole is very similar and in fact gives a larger effect by a factor of two for the modifications to the hyperfine structure, as explained in the Appendix. Under the above set of assumptions, the effect is constrained to be very small (with $\alpha \gtrsim 10^9$ based on the neutrality of the hydrogen atom), and in fact rules out any version of ER = EPR that predicts $\alpha \sim O(1)$. Finally, these constraints have been derived assuming the wormholes are spacelike; more work is needed to extend these results to timelike wormholes as proposed in \cite{Furquan:2025sox}.

Besides the constraints derived above for the hydrogen atom, it is worth pointing out that there exist other avenues to constrain the effect of the electric field of an entangled charged particle entering an ER = EPR wormhole.

First, to obtain even more stringent bounds through the hyperfine structure, it may be helpful to consider other atoms. Since hydrogen is very light, it is relatively hard to trap in comparison to heavier atoms. Hyperfine control (entangling the nucleus and electrons through their spins) is routinely performed for heavier atoms \cite{fernholz2008spin, Veldman:2023rue, Lim:2024sgu}. Heavy Rydberg atoms provide one example, and the spectra of, for example, cesium, rubidium and potassium are also known to a similar high degree of precision as hydrogen \cite{goy1982millimeter, weber1987accurate, li2003millimeter, han2006rb, mack2011measurement, deiglmayr2016precision, peper2019precision}. For this reason, it may be possible for Rydberg atoms to provide even stronger constraints on this effect.

There are also other experiments that could potentially constrain this effect. First, entanglement witness experiments (that have recently been proposed to test the quantum nature of gravity \cite{Bose:2017nin, Marletto:2017kzi, Christodoulou:2018cmk, Christodoulou:2022mkf}) could also be implemented using the electric force to entangle particles. This entanglement would then, according to ER = EPR, generate a quantum wormhole, which, in turn, would have an effect on the electric force between the two particles (according to the assumptions we have made here, if at least one of the particles is a point particle), and this effect could be measured or constrained by the experiment.

Another related experiment would be to put an electron and a charged nanodiamond in traps close to each other such that they interact electrically (see \cite{Poddubny:2024jkq, Winkler:2024fjn} for a similar experiment applied to gravity). In terms of the normal mode coordinates, the ground state is simply a product state, but with respect to the coordinates of the particles, the centers of masses of the electron and nanodiamond are entangled, and the amplitude of this entanglement will depend on how much (if any) of the electric field surrounding the electron goes into the putative ER = EPR wormhole; this provides another path to constrain this potential effect. 

We leave a detailed exploration of these other potential experiments for future work.

\textbf{\textit{Appendix}}---If ER = EPR wormholes are traversable, then (keeping the other assumptions unchanged) there is a similar effect, but with a different amplitude.

If the wormhole is traversable, then the electric field from the electron that passes through the wormhole will exit near the proton, giving the proton an effective charge of $e'' = e + (e'-e) = e/(1+s/\pi\alpha^{2})$, where the second contribution $(e'-e)$ came from the wormhole mouth at the proton (and traveled through the wormhole from the electron). The energy spectrum for the hydrogen atom is then given by
\begin{equation}
\begin{aligned}[b]
E_{n} &= -\frac{m_{e}}{2\hbar^{2}}\left(\frac{e'e''}{4\pi\epsilon_{0}}\right)^{2}\frac{1}{n^{2}}\\
&= -\frac{m_{e}}{2\hbar^{2}}\left(\frac{e^{2}}{4\pi\epsilon_{0}}\right)^{2}\left(1+\frac{s}{\pi\alpha^{2}}\right)^{-4}\frac{1}{n^{2}},
\end{aligned}
\label{eq. 8}
\end{equation}
so the correction, that for large $\alpha$ can be approximated by a factor of $1 - 4 s/ (\pi \alpha^2)$, is twice as large as for non-traversable wormholes.

Similarly, the effect on the hyperfine structure is also twice as strong as for the case of non-traversable ER = EPR wormholes. On the other hand, the total charge in this case remains zero, so entangled hydrogen atoms remain neutral with traversable wormholes (as opposed to the non-traversable case which gave the strong constraint of $\alpha \gtrsim 10^9$).

\textbf{\textit{Acknowledgments}}---The authors would like to thank Marios Christodoulou, Abdullah Irfan, Mohsin Raza, Anton Zasedatelev, and especially Viqar Husain for helpful discussions. This work was supported in part by the Natural Sciences and Engineering Research Council (NSERC) of Canada.

\raggedright

\end{document}